\documentclass[english,superscriptaddress,prl,twocolumn]{revtex4-1}
\usepackage[utf8]{inputenc}
\usepackage[toc,page,title,titletoc,header]{appendix}
\usepackage[T1]{fontenc}
\usepackage{enumerate}
\usepackage{xcolor}
\usepackage{graphicx}
\usepackage{epstopdf}
\usepackage{amsthm}

\usepackage{amsmath}
\usepackage{shapepar}
\usepackage{color}

\usepackage{amssymb, amsbsy, amsthm}
\makeatletter

\newtheorem{Theorem}{Theorem}

\usepackage{babel}
\makeatother

\begin{document}
\title{Uncertainty-Disturbance relations and applications}

\author{Liang-Liang Sun}\email{sun18@ustc.edu.cn}
\affiliation{Hefei National Research Center for Physical Sciences at the Microscale and School of Physical Sciences, University of Science and Technology of China, Hefei 230026, China}

\author{Kishor Bharti}
\affiliation{Centre for Quantum Technologies, National University of Singapore, 3 Science Drive 2, Singapore 117543, Singapore}

\author{Xiang Zhou}
\affiliation{Hefei National Research Center for Physical Sciences at the Microscale and School of Physical Sciences, University of Science and Technology of China, Hefei 230026, China}

\author{ Leong-Chuan Kwek}\email {kwekleongchuan@nus.edu.sg}
\affiliation{ MajuLab, CNRS-UNS-NUS-NTU International Joint Research Unit,UMI 3654, Singapore}
\affiliation{  National Institute of Education, Nanyang Technological University, 1 Nanyang Walk, Singapore 637616}
\affiliation{  School of Electrical and Electronic Engineering Block S2.1, 50 Nanyang Avenue, Singapore 639798}

\author{Jingyun Fan} \email{Fanjy@sustech.edu.cn}	
\affiliation{Shenzhen Institute for Quantum Science and Engineering and Department of Physics, Southern University of Science and Technology, Shenzhen, 518055, China}
\affiliation{Guangdong Provincial Key Laboratory of Quantum Science and Engineering, Southern University of Science and Technology, Shenzhen, 518055, China,}

\affiliation{Hefei National Laboratory, Hefei 230088, China.}

\author{Sixia Yu}\email {yusixia@ustc.edu.cn}
\affiliation{Hefei National Research Center for Physical Sciences at the Microscale and School of Physical Sciences, University of Science and Technology of China, Hefei 230026, China}
\affiliation{Hefei National Laboratory, Hefei 230088, China.}

\date{\today{}}
\begin{abstract}
Uncertainty and intrinsic measurement disturbance, two fundamental concepts in quantum measurement, have conventionally been viewed as distinct and studied separately.  In this work, we establish a fundamental connection between them, proving that uncertainty not only serves as a prerequisite for intrinsic disturbance but also bounds it from above. We formalize this connection via uncertainty-disturbance relations (UDRs) with direct applications in quantum information science. We show that for rank-one projective measurements, these UDRs effectively function as uncertainty relations by bounding the uncertainties of incompatible measurements. They also enable the experimental estimation of key quantum resources---including von Neumann entropy, purity, coherence, and genuine randomness. Our findings thus unify the understanding of uncertainty and disturbance and provide a versatile framework for quantum resource detection.

\end{abstract}
\pacs{98.80.-k, 98.70.Vc}
\maketitle 
%a POVM is a set of positive semi-definite observables that sum to   the identity

\section{I. Introduction} 
Unlike classical measurements—which passively record deterministic observable values—quantum measurements are inherently probabilistic and inevitably perturb the measured quantum state. These two defining features, namely, uncertainty and disturbance, lie at the heart of quantum measurement theory and underpin nearly all quantum phenomena, including nonlocality~\cite{PhysRevA.75.032304, PhysRevA.73.012112, PhysRevA.106.032213}, contextuality~\cite{PhysRevLett.109.050404, RevModPhys.94.045007}, and coherence~\cite{PhysRevA.97.062308, PhysRevA.106.042428}. Moreover, they give rise to fundamental distinctions between classical and quantum information science, serving as key ingredients for quantum advantages in a wide range of information-processing tasks. Given their theoretical and practical significance, a central challenge remains: to refine and expand our understanding of these phenomena.

To date,  uncertainty and disturbance are primarily  associated with the uncertainty principle~\cite{Heisenberg1927,Kennard1927,robertson1929uncertainty,beg,entr,entr1,det,mas, berm, PhysRevA.84.052117,sim,PhysRevLett.111.230401,stronger} and complementary principles~\cite{Ozawa03,ozawa200321,rozema, Werner1,
 	Branciard6742,PhysRevLett.112.020401,PhysRevLett.112.020402,PhysRevLett.112.050401,Werner2,Werner3,
 	PhysRevA.90.042113,PhysRevA.89.022106,PhysRevA.89.052108,busch2006complementarity,RevModPhys.86.1261,
 	PhysRevLett.115.030401,PhysRevLett.114.070402,PhysRevLett.117.140402,EXPW1,EXPW2,Barchielli2018,ylm2019}, respectively. The uncertainty principle states  that a quantum system cannot be  prepared in a state such that incompatible quantities possess exact values simultaneously. This principle has been articulated through various uncertainty relations that utilize different measures, including standard deviation~\cite{Heisenberg1927,Kennard1927,robertson1929uncertainty}, the $\alpha$-R\'{e}nyi entropy~\cite{mas}, and general Schur-concave functions~\cite{PhysRevLett.111.230401}. Moreover, uncertainty relations have been extended to encompass scenarios involving memory~\cite{berm, Xiao2016}, multiple observables~\cite{PhysRevLett.131.150203, PhysRevA.108.042208}, and quantum dynamical processes~\cite{PhysRevResearch.3.023077}. These advancements have proven crucial in establishing the security of quantum communications~\cite{Wootters1982, DIEKS1982271}, verifying quantum randomness~\cite{randomness}, constraining the strength of quantum nonlocality~\cite{Oppenheim2010, PhysRevA.106.032213}, and delving into the foundations of quantum thermodynamics~\cite{PhysRevLett.125.050601}.

The complementary principle asserts that incompatible quantities cannot be precisely measured in  sequential measurements, as expressed by the inherent trade-off between the error of the first measurement and the disturbance imposed on the subsequent one. This principle fundamentally limits the information extractable from a single particle through incompatible measurements~\cite{Park1970}, thereby underpinning the security of quantum cryptographic protocols~\cite{Silva2015, Steffinlongo2022, Miklin2020, Cai2025}. To date, numerous error-disturbance relations have been investigated both theoretically and experimentally~\cite{Ozawa03,ozawa200321,rozema,Werner1,
 	Branciard6742,PhysRevLett.112.020401,PhysRevLett.112.020402,PhysRevLett.112.050401,Werner2,Werner3,
 	PhysRevA.90.042113,PhysRevA.89.022106,PhysRevA.89.052108,busch2006complementarity,RevModPhys.86.1261,
 PhysRevLett.115.030401,PhysRevLett.114.070402,PhysRevLett.117.140402,EXPW1,EXPW2,Barchielli2018,ylm2019}. Nevertheless, a universally accepted definition of intrinsic disturbance in the context of general quantum measurement processes remains elusive~\cite{e21020142}.

The above two kinds of relations treat uncertainty and disturbance as independent properties. 
However, in a quantum measurement process, these two effects manifest simultaneously in a "random jump'' of the measured state, known as the reduction of the wave function, raising the question of whether they are fundamentally interconnected. Winter first observed that, when a measurement is implemented using a L\"uders instrument, the state change induced by the measurement can be upper-bounded by the probability of outcomes---a result now known as the gentle measurement lemma. Although this lemma is specific to L\"uders instruments, it has proven remarkably useful, playing key roles in diverse areas such as:   channel coding theorem~\cite{Winter1999},  state tomography~\cite{PRXQuantum, 10113, Chia20}, self-testing~\cite{PhysRevA.106.L010601},  differential privacy~\cite{10.1145/3313276.3316378}, quantum process learning~\cite{PRXQuantum.5.020367, PhysRevLett.129.160503}, quantum correlation ~\cite{PhysRevX.13.041001, PhysRevLett.129.250504}, $et. al$.    In our previous papers, by analyzing quantum measurement within the framework of Heisenberg's picture~\cite{sun2025}, we show that the uncertainty associated with one measurement can serve as an upper bound for its intrinsic disturbance effect. This finding allowed us to establish an uncertainty-disturbance relation (UDR) and a generalized gentle measurement. Furthermore, we showed that these uncertainty-disturbance relations can be employed to gain insights into the strength of nonlocality in the framework of generalized probabilistic theories~\cite{PhysRevA.106.032213}. 

In this paper, we present a systematic approach to derive uncertainty-disturbance relations, followed by illustrating their applications in quantum information science. We first review basic notions  of a  general measurement. We then introduce the concept of distance-based intrinsic disturbance and show that, for commonly used state-distance measures, the defined intrinsic disturbance induced by a measurement is always upper bounded by its uncertainty and provide many  UDRs.  When restricted to rank-one projection measurements, these UDRs imply constraints on the uncertainties of incompatible measurements, leading to uncertainty relations. We discuss the tightness of these uncertainty relations in the contexts of qubit and qutrit systems. Additionally, UDRs are utilized to design protocols for experimentally estimating key quantities in quantum information science, including the von Neumann entropy, purity, coherence, and quantum Fisher information.

\section{II. Uncertainty-disturbance relation}
We begin with a brief review of the notions of quantum measurement.  
A general quantum measurement is described with a set of positive operator-valued measures (POVMs) $\{{ M}_{i}\}_{i}$.  To realize a non-projective POVM, one has to extend it, using  Naimark's dilation theorem, to be a projective measurement at the cost of introducing a higher-dimensional space. This extension is often described as~\cite{niel}:  coupling the measured   state $\rho_{s}$ with an ancillary state $\rho_a=| 0\rangle \langle 0|_{a}$ via unitary operation $\mathrm{U}$, and performing projective measurement $\{\mathrm{P}_i=|i\rangle \langle i |_{a} \otimes \openone_{s}\}$ on the resulted  joint state  ${\rm U}\rho_{s}\otimes \rho_{a}{\rm U}^{\dagger}$, where $\{| i \rangle \langle i |_{a}\}$ is a set of orthogonal basis acting on the ancillary system. Then,  upon outcome $i$ 
  the measured state  is updated  (up to a normalization) according to the rule  ${\mathcal{J}}^{\mathcal{D}}_{i}:\rho_{s} \to {\rm U}_{i} {\sqrt{\rm M}_{i}}\rho_{s}  {\sqrt{\rm M}_{i}}{\rm U}_{i}$, where ${\rm U}_{i}$ is  determined by ${\rm U}$ and can be arbitrary. This channel does not well characterize the intrinsic disturbance in the measurement process, as the channels $\{\mathcal{J}^{\mathcal{D}}_{i}\}_{i}$  include  additional and reversible state changes induced by ${\rm U}_{i}$. In contrast, intrinsic disturbance stems from solely the irreversible wave function reduction. To rigorously characterize intrinsic disturbance, we adopt the Heisenberg picture.   The POVM is equivalently  realized  by performing $\{{\rm Q}_{i}={\rm U}^{\dagger}{\rm P}_{i}{\rm U}\}$ directly on $\rho_{s}\otimes \rho_{a}$ as   ${\rm Tr}(\rho_{a}\otimes \rho_{s}{\rm Q}_{i})={\rm Tr}(\rho_{a}\otimes \rho_{s} {\rm U}^{\dagger}{\rm P}_{i}{\rm U})={\rm Tr}({\rm U}\rho_{a}\otimes \rho_{s} {\rm U}^{\dagger}{\rm P}_{i})$.  Then, one has a different state updating rule 
  \begin{eqnarray}
{\mathcal{J}^{\mathcal{N}}_{i}}: \rho_{s}\to {\rm Tr}_{a}({\rm Q}_{i} \rho_{s}\otimes \rho_{a}{\rm Q}_{i}),
\end{eqnarray}
where the partial trace is taken over the ancillary system.   {This channel actually characterizes the intrinsic disturbance in a general measurement process. Note that,  an arbitrary   measurement   channel ${\mathcal{J}}_{i}(\cdot)$ can be formulated  with some  projective measurement $\bf Q$, a dilation of the given measurement, on the composite system $\rho_{s}\otimes\rho_{a}$ followed by  unitary operations $\{{\rm U}_{i}\}$  \cite{bush2016} that possibly  depends on the  outcome $i$: $\mathcal {J}_{i}(\rho_s)= 
{\rm Tr}_{a}({\rm U}_{i}{\rm  {Q}}_{i}\rho_{s} \otimes \rho_{a}{\rm  {Q}}_{i} {\rm U}_{i}^{\dagger})$, where the intrinsic disturbance  in the joint system comes from $\bf Q$  ( ${\rm U}_{i}$ introduces a reversible state change).   We refer Ref.\cite{sun2025} for a detailed discussion of this channel. 
Intrinsic disturbance relevant to ${\bf Q}$  can be uncontroversially  defined in terms of the  distance $\mathcal{D}$ between the  initial state  $\rho_{sa}\equiv \rho_{s}\otimes \rho_{a}$  and the post-measurement state $\rho'_{sa}\equiv \sum_{i}{\rm Q}_{i}\rho_{s}\otimes \rho_{a}{\rm Q}_{i}$ as}
\begin{eqnarray}
\mathcal{D}_{\rm D}(\rho_{sa})\equiv{\rm D}(\rho_{sa}, \rho'_{sa}).\label{disdej}
\end{eqnarray}
 where state distance measure ${\rm{D}}(\cdot, \cdot )$ assumes a few nice  properties, namely,  positivity,   unitary invariance, and monotonicity under physical operations.
By tracing out the ancillary system, we have the disturbance in  the measured system $\mathcal{D}_{\rm D}(\rho_{s})$
\begin{eqnarray}
\mathcal{D}_{\rm D}(\rho_{s})\equiv{\rm D}(\rho_{s}, \rho'_{s}), \label{disde}
\end{eqnarray}
 where $\rho'_{s}\equiv\sum_{i}\mathcal{J}^{\mathcal{N}}_{i}(\rho_{s})$ and  our first result is, 
 \begin{Theorem}
[Uncertainty upper-bounds disturbance]
 For a general  state-distance based disturbance  $\mathcal{D}_{\rm D}(\rho_{s})$ or $\mathcal{D}_{\rm D}(\rho_{sa})$, there exists  at least one distance-dependent uncertainty measure $\delta_{D}(\mathbf{p})$, as listed in Table I, serving  as an upper bound 
 \begin{equation}
\begin{aligned}
\delta_{\rm{D}}(\mathbf{p})\geq \mathcal{D}_{\rm D}(\rho_{sa})\geq  \mathcal{D}_{\rm D}(\rho_{s}).   \label{equd}
 \end{aligned}
\end{equation}
where $\mathbf{p}\equiv \{p_{i}={\rm Tr}(\rho_{s}{\rm M}_{i})\}$ specifies the distribution of the measurement of interest.  
\end{Theorem}
 In Table I,  we give the uncertainty relevant to almost all the frequently used state distance measures (see the appendix for the proof), including  the infidelity (If) measure, the trace-distance (Tr)~\cite{niel},  Tsallis relative entropy (Ts)~\cite{tis}, the R\'{e}nyi divergence (Rd) measure~\cite{div}, the relative entropy measure (Re), and the Hilbert-Schimdt (HS) distance measure~\cite{hsd}\footnote{Hilbert-Schimdt distance is not monotonic under general physical operation while monotonic under projection measurements. For this distance, the UDR is established for projection measurement}.   Concerning the results provided in Table I,  we note that some state distances are not symmetric, namely, ${\rm D}(\rho_{1}, \rho_{2})\neq{\rm D}(\rho_{2}, \rho_{1})$,  which may induce different UDRs, and an example is the R\'{e}nyi divergence (see appendix).   {In the appendix, we also establish Winter's gentle measurement lemma for $\mathcal{J}^{\mathcal N}_{i}$, a result originally developed for L\"uders' instrument.  }

\begin{table}[t]
\centering
 \begin{tabular}{lll}
\hline\hline
State distance & $\mathcal{D}(\rho_{1}, \rho_{2})$ & Uncertainty \\
\hline
$\rm{IF}$ & $\sqrt{1 - \mathrm{F}^{2}(\rho_1,\rho_2)}$ & $\delta_{\operatorname{Tr}}(\mathbf{p})$ \\
${\rm D}_{\operatorname{Tr}}$ & $\frac{1}{2}\operatorname{Tr}|\rho_{1} - \rho_{2}|$ & $\delta_{\operatorname{Tr}}(\mathbf{p})$ \\
${\rm D}^{\alpha}_{{\rm Ts}}$ & $\frac{1}{\bar{\alpha}}\left[1 - \operatorname{tr}(\rho^{\alpha}_{1}\rho^{\bar{\alpha}}_{2})\right]$ & $\frac{1 - \|\mathbf{p}\|_{2-\alpha}^{2-\alpha}}{\bar{\alpha}}$ \\
${\rm D}^{\alpha}_{{\rm Rd}}$ & $\frac{-1}{\bar{\alpha}} \log \operatorname{tr}\left[ \left( \rho^{\frac{\bar{\alpha}}{2\alpha}}_{2} \rho_{1} \rho^{\frac{\bar{\alpha}}{2\alpha}}_{2} \right)^{\alpha} \right]$ & $\mathrm{H}_{\frac{1}{\alpha}}(\mathbf{p})$ \\
$S(\cdot \|\cdot)$ & $\operatorname{Tr}[\rho_{1}(\log\rho_{1} - \log\rho_{2})]$ & ${\rm H}(\mathbf{p}) - S(\rho) $ \\
${\rm D}_{\rm HS}$ & $\sqrt{\operatorname{Tr}((\rho_{1} - \rho_{2})^{2})}$ & $\sqrt{\operatorname{Tr}(\rho^{2}) - \|\mathbf{p}\|^{2}_{2}}$ \\
\hline\hline
\end{tabular}
  \caption{Uncertainty measures  induced by distance-based disturbance. We have used notions   ${\rm IF}(\rho_{1}, \rho_{2})\equiv 
 \sqrt{1-\rm{F}^{2}(\rho_1,\rho_2)}$ with $\rm{F}(\rho_1,\rho_2)=[\rm{tr}
\large(\textstyle\sqrt{\rm{\sqrt{\rho_{2}}\rho_{1}\sqrt{\rho_{2}}}}\large)]^{2}$,     $\delta_{\rm Tr}(\mathbf{p})\equiv\sqrt{1-\|\mathbf{p}\|^{2}_{2}}$, whose square is the so-called operationally invariant information~\cite{PhysRevLett.83.3354},  the notions $\bar \alpha \equiv1-\alpha$,  $\rm{H}_{\alpha}(\mathbf{p})\equiv\frac{\alpha}{\bar\alpha}\log \|\mathbf{p}\|_{ \alpha}$ for   R\'{e}nyi entropy, the von von Neumann $S(\rho)\equiv-{\rm Tr}(\rho\log_{2}\rho)$ entropy, and  the Shannon entropy reads  ${\rm H}(\mathbf{p})\equiv-\sum_{i}p_{i}\log_{2} p_{i}$  and consider the regions  $0\leq\alpha< 1$ for R\'{e}nyi divergence and $0\leq\alpha<1$ for  Tsallis relative entropy.}\end{table}

%\subsection{C. On the disturbance in the subsequent  measurement}
Next, we consider the disturbance in a sequential measurement scheme where another following measurement  $B\equiv\{M_{j|B}\}$  is performed on the post-measurement state  $\rho'_{s}$ after measurement $A$,  giving a disturbed statistics  $\mathbf{q}'\equiv\{q'_{j}={\rm Tr}(M_{j|B} {\rho}'_{s} )\}$. Compared probability distribution $\mathbf{q}\equiv\{q_{j}={\rm Tr}(\rho_{s} M_{j|B})\}$ from measurement performed on the original state,  a disturbance in measurement $B$ can be defined with the distance between  $\mathbf{q}'$  and  $\mathbf{q}$.  According to the  data processing inequality, ${\rm{D}}(\rho_{s}, {\rho}'_{s}) \geq \tilde {\rm D}(\mathbf{q}, \mathbf{q}'), $
where  $\tilde {\rm D}(\mathbf{q}, \mathbf{q}')$ is the classical counterpart of state distance ${\rm D}$.   \begin{equation}
\begin{aligned}
 \delta_{\rm{D}}(\mathbf{p})\geq   {\rm D}(\mathbf{q}, \mathbf{q}').\label{UDRm}
 \end{aligned}
\end{equation}
This relation is also called UDR as it is an immediate consequence of Theorem 1. We also give this kind of UDR for each disturbance in Appendix.  In what follows, we demonstrate that UDRs can provide a systematic framework for deriving uncertainty  relations when measurements $A$ and $B$ are rank-one projective measurements.

 {Here, we consider the tightness of these  UDRs. In the derivation of $\delta_{\mathrm{D}} ({\bf  p}) \geq {\rm D}(\rho_{s}, \rho'_s)$,  we have  purified  the joint  state  $\rho_{s}\otimes \rho_{a}$ by introducing environment and then tracing out both the ancilla and the environment. This generally relaxes the inequality. However, when measurement  $A $  is projective measurements and  the measured state is pure, this relaxation can be avoided, and the inequality  for IF, ${\rm{D}}^{\alpha}_{{\rm Rd}}$, $S$, $\operatorname{D}^{\alpha}_{\rm Ts}$, ${{\rm D}_{\rm{HS}}}$, and ${\rm D}_{\rm Tr}$ can be saturated. Furthermore, in  relation  \ref{UDRm} we have  relaxed  ${\rm D}(\rho, \rho_{s})$ to ${\rm D}({\bf q}, {\bf q'})$. For  $\rm IF$ and ${\rm D}_{\rm HS}$,  the 
 relation ${\rm D}(\rho, \rho_{s})\geq {\rm D}({\bf q}, {\bf q'})$ can be saturated when $B$ is an eigenvector of $\rho' - \rho'_s$. To gain further insight on the tightness, when restricted to rank-one projective measurements, these UDRs can be viewed as standard uncertainty relations. We compare their tightness with the Maassen-Uffink uncertainty relation and the relation associated with Winter's gentle measurement lemma (after summing  all the outcomes).  A summary of these UDRs is provided in Table~II. 
}

 \section{III.  UDRs  as preparation uncertainty relations}

Consider sequential rank-one projective measurements \( A = \{ |A_i\rangle \langle A_i| \} \) and \( B = \{ |B_j\rangle \langle B_j| \} \) performed in the order \( A \to B \) on a quantum state \( \rho \) (here $\rho_{s}$ is written as \( \rho \) for brevity). After measuring \( A \), the post-measurement state becomes $\rho' = \sum_i p_i |A_i\rangle \langle A_i|,$ where \( p_i = \mathrm{Tr}(\rho |A_i\rangle \langle A_i|) \). A subsequent measurement of \( B \) on \( \rho' \) yields the statistics \( \mathbf{q}' \equiv \{ q'_j = \sum_i p_i c_{ij} \} \), where the matrix \( \mathcal{C} \equiv \{ c_{mn} = |\langle A_m | B_n \rangle|^2 \} \) encodes the squared overlaps between the eigenvectors of \( A \) and \( B \). This matrix \( \mathcal{C} \) is a unistochastic matrix~\cite{uni1, uni2}. We observe that the disturbed statistics \( \mathbf{q}' \) are simply a linear transformation of the original probabilities \( \mathbf{p} \), meaning that all UDRs listed in Table I are expressed in terms of \( \mathbf{p} \), \( \mathbf{q} \), and \( \mathcal{C} \), and UDRs impose constraints on the allowed probability distributions for a given overlap matrix \( \mathcal{C} \) thus function as uncertainty relations.

\begin{Theorem}[UDRs as uncertainty relations]
When considering rank-one projective measurements, UDRs constrain the probability distributions from incompatible and rank-one measurements on identically prepared ensembles, thereby serving as uncertainty relations in this context.
\end{Theorem}
As uncertainty relations, UDRs are specified with a subscript notation. For example, $\mathrm{U}_{\mathrm{Re}}$ represents the relation:
\[
\mathrm{H}(\mathbf{p}) \geq \mathrm{H}(\mathbf{q}\|\mathbf{q}'),
\]
where $\mathrm{H}(\mathbf{p}) = -\sum_{i} p_{i}\log_{2} p_{i}$ is the Shannon entropy and $\mathrm{H}(\mathbf{q}\|\mathbf{q}') = \sum_{i} q_{i}\log_{2} \frac{q_{i}}{q'_{i}}$ is the classical relative entropy. This corresponds to an uncertainty relation based on the relative entropy distance measure. From $\mathrm{U}_{\mathrm{Re}}$, we can derive the Maassen-Uffink (MU) uncertainty relation~\cite{mas}:
\[
\mathrm{H}(\mathbf{p}) + \mathrm{H}(\mathbf{q}) \geq -\log_{2} c,
\]
where $c \equiv \max_{ij} c_{ij}$. The derivation follows from:
$\mathrm{H}(\mathbf{q}\|\mathbf{q}') \geq \sum_{i} q_{i}\log_{2} \frac{q_{i}}{q'_{i}} \geq -\mathrm{H}(\mathbf{q}) - \sum_{i} q_{i}\log_{2}q'_{i} \geq -\mathrm{H}(\mathbf{q}) - \log_{2} c,$ where we have used $q'_{i} = \sum_{j} c_{ij}p_{j} \leq c$. Combining this with $\mathrm{H}(\mathbf{p}) \geq \mathrm{H}(\mathbf{q}\|\mathbf{q}')$ yields the MU relation. Similarly, from $\mathrm{U}_{\rm ts}$ we can derive an independent entropic uncertainty relation:
\begin{equation}
\frac{1}{2-\alpha}\mathrm{H}_{2-\alpha}(\mathbf{q}) + \mathrm{H}_{\alpha}(\mathbf{p}) \geq -\log c, \quad 0 \leq \alpha < 1, \label{neun}
\end{equation}
which represents a distinct family of relations beyond the Maassen-Uffink framework~\cite{mas}.

Furthermore, in previous uncertainty relations, e.g., the MU entropic uncertainty relation,  the incompatibility of measurements is characterized by a single number $c$, namely, the maximum element of the matrix $\mathcal{C}$. In contrast,  UDRs involve the whole matrix of $\mathcal{C}$. 
Hence, these uncertainty relation arising from UDRs may provide tighter  constraints than the MU entropic uncertainty relations. Below we present two case studies of the \emph{uncertainty-disturbance} type uncertainty relation  regarding  the tightness in the case of qubit ($d=2$) and qutrit ($d=3$).

\begin{table}[t]
	\centering
	\begin{tabular}{lcc}
		\hline\hline
		Relations& Volume(d=2) & Volume(d=3) \\
\hline
$\rm{U}_{\rm Tr}$ &  0.930 & 0.94675 \\
$\rm{U}^{0.5}_{\rm Rd}$ &  0.787 & 0.917 \\
 $\rm{U}_{\rm Re}$ & 0.770 &   0.905 \\
 $\rm{U}^{0.5}_{\rm Ts}$& 0.814 & 0.937   \\
$\rm{U}_{\rm HS}$ &  0.705 & 0.887 \\
${\rm \textstyle{MU}}$ & 0.974&   0.999 \\
        \hline\hline
	\end{tabular}
\caption{Computed \emph{volumes} according to different  UDRs and  their duals.  }
\end{table}

For the qubit case, $\mathbf{p}$, $\mathbf{q}$, and $\mathcal{C}$ are determined by three independent parameters, $\{p_{0},q_{0}, c_{00}\}\in[0,1]$.  We shall investigate constraints placed by various uncertainty relations to $p_{0}$, $q_{0}$, and $c_{00}$. Geometrically one may visualize that each one of the inequalities in   $\rm{U}_{\rm tr}$,   $\rm{U}^{\alpha}_{\rm ts}$, $\rm{U}^{\alpha}_{\rm rd}$,  $\rm{U}_{\rm re}$ and the MU relation for measurements $A\to B$ together with their dual inequalities for measurements $B\to A$  enclose a region of parameter space shown in Fig.1.

We also compute  their \emph{volumes} of the enclosed region in Table II. To this end,   {we randomly select \(10^7\) triples \(\{p_0, q_0, c_{00}\}\) and count the number \(n\) of them that satisfy the uncertainty relation. The \emph{volume} is then estimated as \(n \times 10^{-7}\). This value provides a direct measure for comparing the tightness of different relations: a smaller \emph{volume} indicates a tighter constraint.}  The volume  is normalized to 1 if without any other constraints. Under constraint of UDRs,  the volumes are   in the range [0.705, 0.930],   which are significantly smaller than the \emph{volume}(=0.974) for the MU relation.  Here, we note tight uncertainty relation is $\langle A\rangle^{2} +\langle B \rangle^{2}+C^{2}-2C\langle A\rangle\langle B\rangle\leq 1$
where $\langle A\rangle=2p_{0}-1$,  $\langle B\rangle=2q_{0}-1$, and $C=c_{00}-c_{01}=2c_{00}-1$, for which,  the volume under exact  quantum constraint~\cite{Li2015} is $0.612$. For the qutrit case,  the UDRs enclose regions with volumes in the range  [0.887, 0.947], which are also significantly smaller than the \emph{volume}(0.999) for the MU relation (see Table II).

 {Uncertainty and disturbance, as two fundamental concepts in quantum measurement, have been  investigated through a variety of relations. Among these, the uncertainty relations~\cite{Heisenberg1927,Kennard1927,robertson1929uncertainty,beg,entr,entr1,det,mas,berm,PhysRevA.84.052117,sim,PhysRevLett.111.230401,stronger} and the error-disturbance relations~\cite{Ozawa03,ozawa200321,rozema,Werner1,Branciard6742,PhysRevLett.112.020401,PhysRevLett.112.020402,PhysRevLett.112.050401,Werner2,Werner3,PhysRevA.90.042113,PhysRevA.89.022106,PhysRevA.89.052108,busch2006complementarity,RevModPhys.86.1261,PhysRevLett.115.030401,PhysRevLett.114.070402,PhysRevLett.117.140402,EXPW1,EXPW2,Barchielli2018,ylm2019} are the most studied, while Winter's gentle measurement lemma offers a distinct perspective by directly linking a measurement's uncertainty and its  disturbance.  The lemma is specific to L\"uders instruments and disturbance quantified by fidelity and trace-distance. Here, we develop the lemma and present a  method applicable to general measurement processes and nearly all state distance measures, with which many variants of UDRs are given. These UDRs, while differing from  uncertainty relations and error-disturbance relations, can  be related to them in specific cases. On one hand, they   can be interpreted as uncertainty relations when the measurements are rank-one and projective. However, when  measurement \(A\) is not rank-one, the disturbed statistics \(\mathbf{q}'\) cannot be expressed solely in terms of the statistics relevant to  measurement $A$  and the overlaps between the measurement elements of \(A\) and \(B\); consequently, the corresponding UDRs are not uncertainty relations.  On the other hand,   certain uncertainty measures employed in UDRs---such as Shannon entropy---allow for an informational interpretation, thus  UDRs reveal  that disturbance lower-bound information gain. This differs conceptually from error-disturbance relations, which   typically provide some \emph{lower} bounds on the disturbance in terms of  the error of the first measurement.   From a technical standpoint, our approach makes full use of the results from quantum information science, including state distance measures and the data processing inequality,  in comparison to   commonly used tools  in prior uncertainty relations, such as the Riesz theorem in entropic uncertainty relations~\cite{mas}, the Schwarz inequality in Heisenberg-type variance-based relations~\cite{Heisenberg1927, Kennard1927, robertson1929uncertainty}, and the underlying majorization-based relations~\cite{PhysRevA.84.052117,PhysRevLett.111.230401}.}

\section{IV. Applications of UDRs in quantum information}

In quantum information science, fundamental quantum properties~\cite{RevModPhys.91.025001}---including purity, randomness, and coherence---play vital roles in enabling key information processing tasks such as quantum cryptography, quantum algorithms, and quantum metrology. Experimentally estimating these resources is essential for assessing a quantum system's capability to perform such tasks. In this section, we demonstrate how UDRs provide an effective framework for addressing this challenging problem.

\begin{figure}
\centering
\includegraphics[scale=0.30]{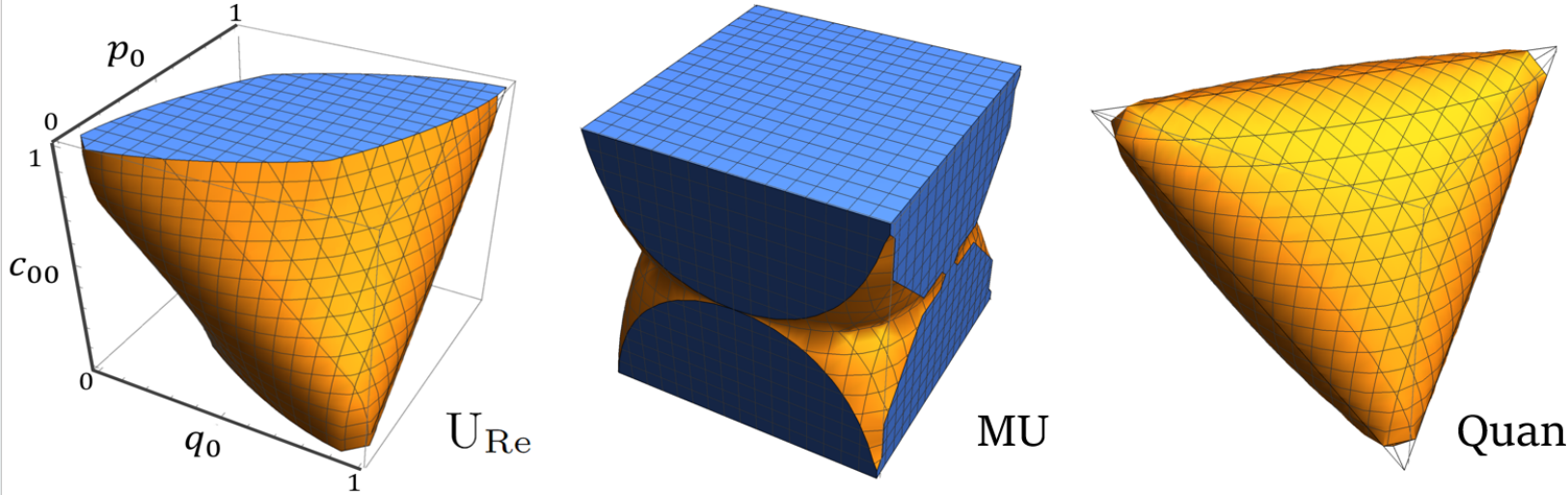}
\caption{Geometrical visualization of the constraints of uncertainty relations. In the qubit case, the distributions from measuring $A$ and $B$ can be specified respectively by $p_{0}$ and $q_{0}$, and their incompatibility is characterized by $c_{00}$. They  are subject to uncertainty relations.  Plots three typical uncertainty relations correspond to   $\rm{U}_{\rm Re}$,  $\rm{U}_{\rm HS}$,   exact quantum uncertainty relation.  } \label{fig:ndwu}  %of
\end{figure}

\emph{The Estimation of von Neumann entropy and purity with UDRs---}
 The first quantity we consider is von Neumann entropy  $S(\rho)=-\rho\log_2 \rho$. It is generally estimated  via the well-known  theory of majorization~\cite{marshall2011inequalities}, where  a projective measurement specified by $A$ performed  on $\rho$  gives a distribution $\mathbf{p}$.  One has $S(\rho)\leq {\rm H}(\mathbf{p}) $.
 However, by the UDR corresponding to the relative entropy, we have
  $$S(\rho)\leq {\rm H}(\mathbf{p})-{\rm H}(\mathbf{q}\|\mathbf{q}'),$$
where the upper  bound is optimized by the disturbance ${\rm H}(\mathbf{q}\|\mathbf{q}')$ measured by the relative entropy between the disturbed (by $A$) and undisturbed statistics of $B$.

Second, we consider the  purity of quantum state, which can be quantified via R\'{e}nyi 2-purity~\cite{GOUR20151} defined as $\log_{2}(d{\rm Tr}(\rho^{2}))$ with $d$ specifying the dimension of the system of interest, $i.e.$, which is a simple function of linear purity ${\rm Tr}(\rho^{2})$.
According to our UDR ${\rm U}_{\rm HS}: \sqrt{{\rm Tr}(\rho^{2})-\|\mathbf{p}\|^{2}_{2}}\geq \sqrt{\|\mathbf{q}-\mathbf{q'}\|^{2}_{2}}$, we have
 \begin{equation}
\begin{aligned}
{\rm Tr}(\rho^{2})\geq \|\mathbf{p}\|^{2}_{2}+\|\mathbf{q}-\mathbf{q}'\|^{2}_{2}. \end{aligned}
\end{equation}
 One can achieve such an estimation with only $\mathbf{p}$,   $\mathbf{q}$ and $\mathbf{q}'$, where  measurement  $B$ needs to be projective.

\emph{Estimation of  coherence and randomness}
In this section, we estimate coherence and randomness measures defined via convex-roof construction~\cite{Konig2009, PhysRevA.92.022124, PhysRevA.96.032316}.
 That is, provided that for a pure state $|\phi\rangle$ the given resource is quantified by some measure, say, $\mathcal{R}(\phi)$. The measure   for a mixed state $\rho$  is  defined via a convex-roof construction as
$$\mathcal{R}(\rho)=\textstyle \min_{\{r_{n},|\phi_{n}\rangle\}}\sum_{n}\mathcal{R}(\phi_{n}),$$
 where the optimization is taken over all the possible pure-state ensembles of $\rho=\sum_{n}r_{n}\cdot|\phi_{n}\rangle\langle \phi_{n}|$. This kind of measure is notoriously difficult to compute and estimate in experiments as there is  infinite possible  decomposition of $\rho$. Our UDRs serve well the issue of estimating convex-roof based resource  for coherence and quantum randoms. This is because, on one hand,  the measures    are often defined as uncertainty or a related function for a pure state.  On the other hand, UDRs have favorable properties:  uncertainty is convex while disturbance is concave. This allows one to convert a convex uncertainty into a concave one. Then we have the following theorem
  \begin{Theorem}[Estimation of convex-roof based single partite resource]
 If a measure for pure state resource $\mathcal{R}(\phi)$  admits a lower bound  in terms of an increasing convex function $f(\cdot)$ of uncertainty $\delta(\mathbf{p})$, i.e., $\mathcal{R}(\phi)\geq f(\delta(\mathbf{p}))$  then we have
\begin{equation}
\begin{aligned}
 \mathcal{R}(\rho)\geq f[  \operatorname{D}(\rho, \rho')]\geq f[  \tilde{\rm {D}}(\mathbf{q}, \mathbf{q'})] , \label{crc}
\end{aligned}
\end{equation}
where $\operatorname {D}(\rho, \rho')$ is state-distance measure corresponding to $\delta(\mathbf{p})$ in an uncertainty-disturbance relation and $\tilde{\rm {D}}(\mathbf{q}, \mathbf{q'})$ can be measured in experiment.
\end{Theorem}

\begin{proof}
Specifying  the optimal decomposition of $\rho$ achieving the convex-roof is $\{r_{n}, |\phi_{n}\rangle \}$
\begin{equation}
\begin{aligned}
\mathcal{R}(\rho)&= \sum_{n} r_{n}\mathcal{R}(\phi_{n})\geq \sum_{n}r_{n}f(\delta(\mathbf{p}_{n}))\geq f( \sum_{n}r_{n} \delta(\mathbf{p}_{n}))\\
&\geq f[\sum_{n}r_{n}{\rm D}(\phi_{n}, \rho'_{n})]
\geq f[{\rm D}(\rho, \rho')]\geq f[\tilde{\rm D}(\mathbf{q}, \mathbf{q'})],
 \end{aligned}
\end{equation}
where $\rho'_{n}$ specify the post-measurement state relevant to $|\phi_{n}\rangle$ and $\rho'=\sum_{n}r_{n}\rho'_{n}$ and the inquality follows the defining  property of $f$ and UDRs.  
  \end{proof}

\emph{Case.1 Convex roof of infidelity.}
Coherence is a property  characterized and quantified  with respect to a computation basis, say, $\{|A_{i}\rangle\}$.
 Consider a convex-roof based  measure~\cite{Liu2017}
 $C_{\rm if}(\rho):=\textstyle \min_{\{r_{n}, |\phi_{n}\rangle \}}\sum_{n} r_{n}\cdot C(\phi_{n}),$
where the measure for a pure state $|\phi\rangle$ is  $ C(\phi)=\sqrt{1-p}$  with  $p\equiv \max_{i}|\langle\phi|A_{i}\rangle|^{2}$.  We immediately have a lower bound
$C_{\rm if}(\rho)\geq\textstyle \frac{\sqrt{2}}{2}D_{\operatorname{tr}}(\rho,  \rho')\geq  D(\mathbf{q}, \mathbf{q}')$ with  $\rho'=\sum_{i}p_{i}|A_{i}\rangle \langle A_{i}|$ (see Appendix).

\emph{Case.2 Quantum Fisher information.}
Quantum Fisher information quantifies the information of one parameter, say $\theta$,  carried in a quantum state $\rho(\theta)$ evolving according to unitary operation $e^{-i A\theta}$. It  determines the ultimate precision of the parameter estimation via the quantum Cram\'{e}r-Rao bound~\cite{Liu2019}.  Quantum fisher information is given as
$F(\rho, A)=\sum_{kl}2\frac{(r_{k}-r_{l})^{2}}{r_{k}+r_{l}}|A_{kl}|^{2}$, where $A_{kl}\equiv\langle \psi_l| A|\psi_k\rangle $  and $r_{i}$ and $|\psi_{i}\rangle $ are the eigenvalues and eigenvectors  of density matrix $\rho$ respectively. By UDRs,  we finally have (see Appendix)
\begin{equation}
\begin{aligned}
F(\rho, A)\geq 4{\rm D}^{2}_{\rm Tr}(\mathbf{q}, \mathbf{q}'),
\end{aligned}
\end{equation}
one can estimate $F(\rho, A)$ via computing  ${\rm D}_{\rm Tr}(\rho, \rho')$  and detect it  via experimentally estimating ${\rm D}_{\rm Tr}(\mathbf{q}, \mathbf{q}')$.

\section{Discussions and Conclusions}
\label{sec:conclusion}
 {Uncertainty and disturbance are two fundamental notions of quantum measurements, classically explored through uncertainty relations and error-disturbance trade-offs. Winter's gentle measurement lemma bridges these concepts in an elegant  light, revealing a direct and insightful connection. In this work, we introduce a systematic framework for deriving such uncertainty-disturbance relations. We demonstrate that the intrinsic disturbance of a general quantum measurement can be bounded by its uncertainty---with each choice of state distance yielding a distinct variant of the relation. When the measurement is rank-one, these UDRs naturally simplify into familiar uncertainty relations. To assess their tightness, we develop a  volume-based approach,  showing that these UDRs imply stronger constraints than the Maassen--Uffink relation for incompatible measurements on qubit and qutrit systems. Furthermore, we apply these UDRs across quantum information science, offering practical protocols for estimating von-Neumann entropy, quantum purity, coherence, and randomness extraction. Thus, our uncertainty-disturbance relations not only deepen the conceptual foundations of quantum mechanics but also open promising pathways for quantum information applications---bringing theory closer to experiment.}

\section{ACKNOWLEDGMENTS}
 L.L.S and  S. Y are  supported by the Quantum Science and Technology-National Science and Technology Major Project (Grant No. 2021ZD0300804). KLC acknowledges funding from Ministry of Education, Singapore and the National Research Foundation, Singapore.  J.Y. F is  supported by the Key-Area Research and Development Program of Guangdong Province Grant No.2020B0303010001, Grant No.2019ZT08X324, No.2019CX01X042, and Innovation Program for Quantum Science and Technology (2021ZD0300804).

\section{DATA AVAILABILITY}

Data sharing is not applicable to this article as no datasets were generated or analyzed during the current study.

\section{Appendix A: Proof of Theorem.1 }
Here, we shall give the proofs for the results in Table I in the main text.  Let the auxiliary system be in a pure state $\rho_{a}$ (as a mixed state can be purified by including an environment)   and specify $\rho_{sae}$ the purification of $\rho_{s}\otimes \rho_{a}$ and  $\rho'_{sae}$ the post-measurement state $\sum_{i}{\rm Q}_{i}\rho_{sae} {\rm Q}_{i}$,  where $\{{\rm Q}_{i}\}$ is the extended measurement realizing the measurement of interest.

\subsubsection{Infidelity}
\emph{uncertainty-disturbance relation}
Provided the measure of   ${\rm IF}(\rho_{1}, \rho_{2})=
 \sqrt{1-\rm{F}^{2}(\rho_1,\rho_2)}$ with $\rm{F}(\rho_1,\rho_2)=[\rm{tr}
\large(\textstyle\sqrt{\rm{\sqrt{\rho_{2}}\rho_{1}\sqrt{\rho_{2}}}}\large)]^{2}$. 
	\begin{eqnarray}
\nonumber {\rm F}(\rho_{s},\rho'_{s})\geq \nonumber {\rm F}(\rho_{sae},\rho'_{sae})&=&\textstyle [{\rm Tr}(\sqrt{ {\sqrt{\rho_{sae}}{\rho}'_{sae}\sqrt{\rho_{sae}}}})]^2\nonumber \\
&=&[{\rm Tr}(\sqrt{ {\rho_{sae}{\rho}'_{sae}\rho_{sae}}})]^2 \nonumber \\
&=&\textstyle [{\rm Tr}(\sqrt{ {\rho_{sae}\sum_{i}{\rm Q}_{i}\rho_{sae} {\rm Q}_{i}\rho_{sae}}})]^2 \nonumber \\
&=&\textstyle [\sqrt{\sum_{i}{\rm Tr}(\textstyle {\rho_{sae}({\rm Q}_{i}\rho_{sae} {\rm Q}_{i})})}]^2 \nonumber \\
&=&\|\mathbf{p}\|^{2}_{2}.
	\end{eqnarray}
 where the first inequality is due to the monotonic property of fidelity under partial trace and the second and the fourth quality  are  due to that the state $\rho_{sae}$ is pure.  
It then follows that $\delta_{\rm IF}(\mathbf{p})\equiv \sqrt{1- \|\mathbf{p}\|^{2}_{2}}$ and $ {\rm IF}(\rho_{s}, \rho'_{s}) =\sqrt{1-{\rm F}^{2}(\rho_{s}, \rho'_{s})}$, then we   have
	\begin{eqnarray}\nonumber
\delta_{\rm IF}\geq \operatorname{IF}(\rho_{s},\rho'_{s})\geq \operatorname{IF}(\mathbf{q},\mathbf{q'}).
	\end{eqnarray}
with  $\operatorname{IF}(\mathbf{q},\mathbf{q'}):=\sum_{i}\sqrt{q_{i}q'_{i}}$ being the infidelity between the disturbed and undisturbed statistics.

\subsubsection{Trace-distance}
For trace-distance ${\rm D}_{\rm Tr}(\rho_{1}, \rho_{2})=\frac12\rm{Tr}|\rho_{1}-\rho_{2}|$, we have proved the corresponding Theorem 1 in the main text. Then we have
\begin{eqnarray}
\textstyle \sqrt{1-\|\mathbf{p}\|_{2}^{2}}&= &\textstyle  \sqrt{1-\sum_{i}p^2_{i}}
 = \sqrt{1-\operatorname{Tr}(\rho_{sae}  {\rho}'_{sae})}\nonumber \\
&\geq& \sqrt{1-\operatorname{F}(\rho_{sae},  {\rho}'_{sae})}
\ge \textstyle \frac{1}{2}\operatorname{tr}|\rho_{sae} - {\rho}'_{sae}|,\nonumber \\
&\geq& \textstyle \frac{1}{2}\operatorname{Tr}|\rho_{s} - {\rho}'_{s}|\nonumber
\end{eqnarray}
where in the second equality we have used the fact that $\operatorname{Tr}(\rho_{sae}  {\rho}'_{sae})=\operatorname{Tr}({\rho}'_{sae})^2=\sum_ip_i^2$   and  in the first inequality we have used the inequality $\operatorname{tr}(\rho_{1} \rho_{2})\leq \rm{F}(\rho_1,\rho_2) $ and the Fuchs-van de Graaf inequality  $\sqrt{1-\rm{F}(\rho_1,\rho_2)}\geq \textstyle \frac{1}{2}{\rm Tr}|\rho_{1}-\rho_{2}|$  in the second inequality. The last inequality is due to the data processing inequality.

\subsubsection{$\alpha-$R\'{e}nyi divergence}
Provided state distance  $\alpha-$R\'{e}nyi divergence ${\rm D }^{\alpha}_{\rm Rd}(\rho_{1}, \rho_{2})= \frac{-1}{\bar \alpha}
\log_2\large\rm{tr}(
\rho^{\frac{\bar \alpha}{2\alpha}}_{2}
\rho_{1}\rho^{\frac{\bar \alpha}{2\alpha}}_{2})^{\alpha}\large$ with $0< \alpha <1$,
we present below the variants of theorem.
As $\alpha-$R\'{e}nyi divergence is not symmetric, i.e.,  ${\rm{D}}^{\alpha}_{\rm Rd}(\rho'_{s}, \rho_{s})$ maybe  not equal to ${\rm{D}}^{\alpha}_{\rm Rd}( \rho_{s}, \rho'_{s})$, we  consider both two cases.

Adhere to the notions $\bar\alpha\equiv1-\alpha$ and $\kappa\equiv \frac{\bar\alpha}{2\alpha}$,  we have
\begin{eqnarray}
{\rm{D}}^\alpha_{\rm Rd}(\rho_{sae}, \rho'_{sae})&=&\textstyle \frac{-1}{\bar\alpha}
\log_{2}\large{\rm{Tr}}(
{\rho}'^{\kappa}_{sae}
{\rho}_{sae}{{\rho}'^{\kappa}_{sae}})^{\alpha}\nonumber \\
&\leq& \textstyle \frac {-1}{\bar\alpha}\log_{2}
	[\operatorname{Tr}({\rho}'^{\kappa}_{sae}
	\rho_{sae}{\rho}'^{\kappa}_{sae})]^{\alpha}
	\nonumber\\
&=&\textstyle \frac {-1}{\bar\alpha} \log_{2}[\operatorname{Tr}
	(\rho_{sae}{\rho}'^{2\kappa}_{sae})]^{\alpha}\nonumber \\
&=&\textstyle \frac {-1}{\bar\alpha} \log_{2}[\sum_{i}p^{2\kappa}_{i}{\rm Tr}(\rho_{sae}\cdot{\rho}'_{i|sae})]^{\alpha}\nonumber\\
&=&\textstyle \frac {-1}{\bar\alpha} \log_{2}[\sum_{i}p^{2\kappa+1}_{i})]^{\alpha}\nonumber\\
&=&\textstyle \frac {-1}{\bar\alpha}\log_{2} (\sum_{i}p^{\frac{1}{\alpha}}_{i})^{\alpha}=
	{\rm H}_{\frac{1}{\alpha}}(\mathbf{p}), \quad \quad\nonumber
	\end{eqnarray}
where   we have used   that, for a positive Hermitian operator $O$, $\operatorname{Tr}O^{\alpha}\geq (\operatorname{Tr}O)^{^\alpha} $ for $ 0< \alpha < 1$ in the second inequality and we have used notation ${\rho}'_{i|sae}={\rm Q}_{i}\rho_{sae}{\rm Q}_{i}/p_{i}$ to specify  the post-measurement state relevant to outcome $i$ that is a pure state and   classical $\alpha$-R\'{e}nyi entropy ${\rm H}_{\alpha}(\mathbf{p})=\frac{{\alpha}}{\bar\alpha}\log _2\|\mathbf{p}\|_{\alpha}$.
Using data processing inequality  we have
\begin{eqnarray}
{\rm H}_{\frac{1}{\alpha}}(\mathbf{p})\geq {\rm{D}}^\alpha_{Rd}(\rho_{sae}, \rho'_{sae}) \geq {\rm{D}}^{\alpha}_{Rd}(\rho_{s}, \rho'_s)\geq \tilde{\rm D}^{\alpha}_{Rd}(\mathbf{q}\|\mathbf{q'}),\nonumber
\end{eqnarray}
where the classical R\'{e}nyi relative entropy $\tilde{\rm D}^{\alpha}_{\rm Rd}(\mathbf{q}\|\mathbf{q'})\equiv \frac{1}{\alpha-1}\log_2 \sum_{i}{q}^{\alpha}_{i}{q'}_{i}^{1-\alpha}$.

Consider the distance ${\rm{D}}^{\alpha}_{{\rm Rd}}(\rho'_{sae}, \rho_{sae})$, we have
\begin{eqnarray}
{\rm{D}}^{\alpha}_{\rm Rd}(\rho'_{sae}, \rho_{sae})&=&\textstyle \frac{-1}{\bar\alpha}
\log_2\large{\rm{Tr}}(
\rho_{sae}^{\kappa}
{\rho}'_{sae}\rho_{sae}^{\kappa})^{\alpha}\nonumber \\
&= &\textstyle\frac {-1}{\bar\alpha}\log_2\large{\rm{Tr}}(
\rho_{sae}
{\rho}'_{sae} \rho_{sae})^{\alpha}\nonumber \\
&=&\textstyle \frac {-1}{\bar\alpha}\log_2 (\sum_{i}p^{2}_{i})^{\alpha}\nonumber\\
&=&\frac {\alpha}{\bar\alpha}{\rm H}_{2}(\mathbf{p}),
\end{eqnarray}
where the second inequality is due to that $\rho_{sae}$ is a pure state.
Using the data processing inequality, we have
\begin{eqnarray}
\frac{\alpha}{\bar\alpha}{\rm H}_{2}(\mathbf{p}) \geq {\rm{D}}^{\alpha}_{\rm Rd}(\rho'_{s}, \rho_s)\geq \tilde{\rm D}^{\alpha}_{\rm Rd}(\mathbf{q'}\|\mathbf{q}).
\end{eqnarray}

\subsubsection{Relative entropy}
\emph{Proof of the variant of Theorem.1  corresponding to  Relative entropy $S(\rho_{1}\|\rho_{2}) =\operatorname{Tr}[\rho_{1}(\log_2{\rho_{1}}-\log_2{\rho_{2}})]$---}. 
 In this case, we only assume $\rho_{a}$ is pure while $\rho_{s}$ is not necessarily a pure state. We have
\begin{equation}
	\begin{aligned} S(\rho_{sa}\|\rho'_{sa}) &={\rm Tr}(\rho_{sa}(\log_2\rho_{sa}-\log_2\rho'_{sa}))\nonumber \\	 
    & \nonumber 
    =-S(\rho_{sa})-\textstyle \sum_{i}p_{i}\log_2 p_{i} +\sum_{i}p_{i}S(\rho_{i|sa}) \nonumber \\
        & \nonumber 
    =-S(\rho_{sa})-\textstyle \sum_{i}p_{i}\log_2 p_{i}  \nonumber \\
   &\leq-S(\rho_{s}\otimes\rho_{a})+\operatorname{H}(\mathbf{p}) \nonumber \\
&=-S(\rho_{s})-S(\rho_{a})+\operatorname{H}(\mathbf{p}) \nonumber \\
&=-S(\rho_{s})+\operatorname{H}(\mathbf{p}).
 \nonumber
	\end{aligned}
	\end{equation}
	where $\rho'_{sa}=\sum_{i}{\rm Q}_{i}\rho_{sa}{\rm Q}_{i}$, ${\rho}_{i|sa}={\rm Q}_{i}\rho_{sa}{\rm Q}_{i}/p_{i}$, and $\rho_{sa}\equiv \sum_{i}p_{i}\rho_{i|sa}$,   and  $-\operatorname{Tr}(\rho_{sa}\log_2\rho'_{sa})=-\sum_{i}p_{i}\log_2 p_{i}+S(\rho_{i|sa})\leq \operatorname{H}(\mathbf{p})$.    Using  the data processing inequality, we have
\begin{eqnarray} \nonumber
	-S(\rho_{s})+\operatorname{H}(\mathbf{p})=S(\rho_{sa}\|\rho'_{sa})\geq  S(\rho_{s}\|\rho'_{s})\geq {\rm H}(\mathbf{q}\|\mathbf{q'}). \nonumber
	\end{eqnarray}

\subsubsection{Tsallis relative entropy}
\emph{Proof of the variant of Theorem.1 corresponding to ${\rm D}^{\alpha}_{\rm Ts}  $---}    Tsallis relative entropy (Ts) ${\rm D}^{\alpha}_{\rm Ts}(\rho_{1}, \rho_{2})=\frac{1}{\bar \alpha}[1-\rm{tr}
(\rho^{\alpha}_{1}\rho^{\bar \alpha}_{2})]$ with $0\leq \alpha<1$ and $\bar \alpha= 1-\alpha$
The distance in terms of Tsallis relative entropy  ($0<\alpha<1$) is given by
	\begin{eqnarray}\nonumber
	\operatorname{D}^{\alpha}_{\rm Ts}(\rho_{sae}, \rho'_{sae}) &=&\textstyle \frac{1}{\bar\alpha}[1-\operatorname{Tr}
	(\rho^{\alpha}_{sae}({{\rho}'_{sae}})^{\bar\alpha})]\nonumber \\
&=&\textstyle \frac{1}{\bar\alpha}[1-\sum_{i}p^{\bar\alpha}_{i}\operatorname{Tr} (\rho_{sae}\cdot{\rho}'_{i|sae})],\nonumber \\
&=&\frac{1- \sum_{i} p_{i}^{2-\alpha}}{\bar\alpha}=\frac{1-\|\mathbf{p}\|_{2-\alpha}^{2-\alpha}}{\bar\alpha}.
	\end{eqnarray}

Considering the disturbance in the following measurement $B$,
\begin{eqnarray}\nonumber
\frac{1-\|\mathbf{p}\|_{2-\alpha}^{2-\alpha}}{\bar\alpha}&=& \operatorname{D}^{\alpha}_{\rm Ts}(\rho_{sae}, \rho'_{sae}) \geq \operatorname{D}^{\alpha}_{\rm Ts}(\rho_{s}, \rho^{(r)}_{s})\nonumber \\
&\geq& {\tilde{\rm D}}^{\alpha}_{\rm Ts}(\mathbf{q}, \mathbf{q'}),
	\end{eqnarray}
which leads to  a neat expression of UDR as
\begin{eqnarray}
	\operatorname{H}_{2-\alpha}(\mathbf{p}) &\geq \tilde{\operatorname{D}}^{\alpha}_{\rm Rd}(\mathbf{q}\|\mathbf{q'}).
	\end{eqnarray}

\subsubsection{Hilbert-Schmidt distance}
Let the state distance be defined as \( \rm D_{\rm HS}(\rho_{1} - \rho_{2}) = \sqrt{{\rm Tr}((\rho_{1} - \rho_{2})^{2})} \). This measure is not monotonic under general physical operations but is monotonic under projective measurements and partial-trace
\begin{eqnarray}
\begin{aligned}
{{\rm D}_{\rm{HS}}(\rho_{sae},\rho'_{sae})}&=\sqrt{{\rm Tr}(\rho_{sae}-{\rho}'_{sae})^{2}}\\ \nonumber
&=\sqrt{{\rm Tr}(\rho^{2}_{sae})+{\rm Tr}({\rho'_{sae}})^{2}-2{\rm Tr}(\rho_{sae}\cdot {\rho}'_{sae})}\\ \nonumber
&=\textstyle \sqrt{{\rm tr}\rho^{2}_{s}-\|\mathbf{p} \|^2_{2}}\nonumber\\
& \ge{\rm{D}}_{\rm{HS}}(\rho_{s},\rho'_s)\geq \tilde{\rm{D}}_{\rm{HS}}(\mathbf{q},\mathbf{q}'). \label{hs}
\end{aligned}
\end{eqnarray}
where $\mathbf{q}$ and $\mathbf{q}'$ are statistics relevant to a projection measurement respectively performed on $\rho_{s}$ and $\rho'_{s}$.

\subsection{Recovering gentle  measurement Lemma}

GML considers L\"{u}ders' instrument that updates a measured state with respect to outcome $i$ into $\rho_{i|\mathcal{L}}=\sqrt{M_{i}}\rho_{s}\sqrt{M_{i}}/p_{i}$ and states
 \begin{eqnarray}
F(\rho, \rho_{i|\mathcal{L}})\geq p_{i}.
\end{eqnarray}
Consider the fidelity between  initial state $\rho_{s}$ and the $i-$th post-measurement state,  we have
 \begin{eqnarray}
{F}(\rho_{s},\mathcal{J}_{i}(\rho))\geq {F}(\rho_{sae},\rho_{i|sae})={\rm Tr}(\rho_{sae}\rho_{i|sae})=p_{i}, 
\end{eqnarray}
where  $\rho_{i|sae}={\rm Q}_{i}\rho_{sae} {\rm Q}_{i}$.

\section{Appendix. B:  The estimate of quantum coherence  and quantum Fisher information}
Consider a convex-roof based coherence   measure
 $C_{\rm if}(\rho):=\textstyle \min_{\{r_{n}, |\phi_{n}\rangle \}}\sum_{n} r_{n}\cdot C(\phi_{n}),$
where the measure for a pure state $|\phi\rangle$ is  $ C(\phi)=\sqrt{1-p}$  with  $p\equiv \max_{i}|\langle\phi|A_{i}\rangle|^{2}$.
As $\sqrt{2}C(\phi)\geq \sqrt{(1-p)(1+p)}\ge \sqrt{1-\|\mathbf{p}\|^{2}_{2}}=\delta_{\rm Tr}(\mathbf{p})$. From Table~I, we have different choices of  UDRs. For an instance, if we take the UDR for trace distance $\sqrt{1-\|\mathbf{p}\|^{2}_{2}}\geq \frac{1}{2}\operatorname{tr}|\rho_{s}-\rho'_{s}|$ with $\rho'_{s}=\sum_{i}p_{i}|A_{i}\rangle\langle A_{i}|$,   we immediately have a lower bound
$C_{\rm if}(\rho)\geq\textstyle \frac{\sqrt{2}}{4}\operatorname{tr}|\rho_{s}- \rho'_{s}|, $
 where the lower bound  does not involve any minimization process  thus is easy to calculate.

\emph{Case.2 Quantum fisher information}
Quantum fisher information quantifies the information of one parameter, say $\theta$,  carried in a quantum state $\rho(\theta)$ evolving according to unitary operation $e^{-i A\theta}$. It  determines the ultimate precision of the parameter estimation via the quantum Cram\'{e}r-Rao bound.  Quantum fisher information is given as
\begin{equation}
\begin{aligned}
F(\rho, A)=\sum_{kl}2\frac{(r_{k}-r_{l})^{2}}{r_{k}+r_{l}}|A_{kl}|^{2},
\end{aligned}
\end{equation}
where $A_{kl}\equiv\langle \psi_l| A|\psi_k\rangle $  and $r_{i}$ and $|\psi_{i}\rangle $ are the eigenvalues and eigenvectors  of density matrix $\rho$ respectively. It is found that $F(\rho, A)$ actually is a convex-roof of variance, namely,
\begin{equation}
\begin{aligned}
F(\rho, A)= 4\min_{\{r_{i}, |\psi_{i}\rangle\}}\sum_{i} r_{i}V(|\psi_{i}\rangle, A),
\end{aligned}
\end{equation}
where $V(\rho, A)={\rm tr}(\rho\cdot  A^{2})- {\rm tr}(\rho\cdot  A)^{2}$.
The minimization is taken over any possible  decomposition of $\rho=\sum_{i}r_{i}\cdot |\psi_{i}\rangle \langle \psi_{i}|$.  In the case that $A$ is binary observable with outcome $\pm 1$, we have $V(\rho, A)=2(1-\|\mathbf{p}\|^{2}_{2})$. By UDRs,  we finally have
\begin{equation}
\begin{aligned}
F(\rho, A)\geq 4 \delta^{2}_{\rm Tr}(\mathbf{p})\geq 4{\rm D}^{2}_{\rm Tr}(\rho, \rho') \geq 4{\rm D}^{2}_{\rm Tr}(\mathbf{q}, \mathbf{q}'),
\end{aligned}
\end{equation}
one can estimate $F(\rho, A)$ via computing  ${\rm D}_{\rm Tr}(\rho, \rho')$  and detect it  via experimentally estimating ${\rm D}_{\rm Tr}(\mathbf{q}, \mathbf{q}')$.

%In this paper, we explore the profound connection between these two quantities by presenting a set of uncertainty-disturbance relations.  These relations illustrate that the uncertainty associated with a measurement serves as a prerequisite for its intrinsic disturbance. While distinct, our findings align well with both traditional uncertainty relations and error-disturbance relations, enriching the discourse on uncertainty principles from a fresh perspective. Our results suggest that intrinsic disturbance is indicative of information gain, thereby complementing the established notion that information gain leads to disturbance, as captured by error-disturbance relations. Furthermore, 

\bibliographystyle{unsrt} 
\bibliography{uncertainty}

\end{document}